\documentclass{llncs}

\usepackage{graphicx}
\usepackage{mathpartir}
\usepackage{prooftree}
\usepackage{verbatim}
\usepackage{color}
\usepackage{tikz}
\usepackage{multirow}
\usetikzlibrary{calc}

\begin{document}

\makeatletter
\newcommand{\thickhline}{%
    \noalign {\ifnum 0=`}\fi \hrule height 1pt
    \futurelet \reserved@a \@xhline
}
\makeatother

\newcommand{\version}{1.5}
\newcommand{\previousversion}{1.4.3}
\newcommand{\reCheck}[1]{{\color{red}#1}}
\newcommand{\ignore}[1]{}

\title{Update report: LEO-II version \version}
\titlerunning{LEO-II}

\newcommand{\ChoiceFuns}{\textsf{CFs}}
\newcommand{\true}{{\mathrm{t}\!\mathrm{t}}}
\newcommand{\false}{{\mathrm{f}\!\mathrm{f}}}
\newcommand{\Leo}{\textsc{Leo}-II}
\newcommand{\rl}[1]{\textsf{#1}} 
\newcommand{\unify}[2]{#1 \neq^? #2} 

\author{Christoph Benzm\"uller\inst{1} \and Nik Sultana\inst{2}}
\authorrunning{Christoph Benzm\"uller \and Nik Sultana}
\tocauthor{}
\institute{Freie Universit\"at Berlin, Germany \and Cambridge University, UK}

\maketitle

\begin{abstract}
Recent improvements of the LEO-II theorem prover are presented. These
improvements include a revised ATP interface, new translations into
first-order logic, rule support for the axiom of choice, detection of
defined equality, and more flexible strategy scheduling.
\end{abstract}

\section{Introduction}
It has been five years since the last system description of \Leo{}
\cite{C26}, and during the last months various improvements have been
made to the system. In this article we outline the current system and
describe the recent improvements.

\section{System overview}
\Leo{} is written in OCaml and implements a RUE calculus \cite{W47}
which relies on a `Boolean aware' (or, more generally, `theory
aware'~\cite{J5}) extensional preunification engine.  \Leo{} accepts
problems encoded in the CNF (clausal first-order form) and FOF
(first-order form) languages from the TPTP \cite{sutcliffe2009tptp},
but its principal input language is
THF0, core typed higher-order form~\cite{sutcliffe2010automated}.

The logical organisation of the prover is illustrated in
Figure~\ref{fig:architecture}, and this roughly corresponds to the
modular organisation of the code. It is structured into four
layers, as the figure shows:
\begin{description}
\item[\emph{Operating mode.}] The prover can be operated in two ways:
    (i) \Leo{} can be used as a proof assistant when run in
    \emph{interactive mode}. It provides a command interface through
    which the user can inspect and manipulate the prover's state, making
    calls to the calculus' rules as needed. This mode is very valuable
    for exploring logical problems and for debugging the prover's
    automatic mode.
    (ii) The prover is usually run in \emph{automatic mode}: this
    comprises a set of strategy schedules, and a main loop which
    drives applications of the calculus' rules.

\item[\emph{Prover interface.}] Both modes use a common infrastructure:
  they parse a problem and load it into the prover's state, then
  further manipulate the state by executing \emph{commands}. A command
  might involve carrying out an inference, inspecting the state,
  switching flags, calling external provers, etc. Each command makes
  calls to lower levels of the prover.

\item[\emph{Logic.}] The main component in this level consists of the
  calculus: a collection of functions which accept and return clauses.
  This level also contains \Leo{}'s main loop, and an interface to
  external ATPs (which also translates problems to other formats).

\item[\emph{Basis.}] The lowest level of \Leo\ defines the
  representation of terms and types, and associated operations
  (e.g. substitution, unification, matching, etc).

\end{description}

\tikzstyle{block} = [draw, fill=gray!10, rectangle, minimum height=2em, minimum width=6em]
\tikzstyle{pill} = [draw, fill=red!60, rounded corners,circle]
\tikzstyle{group} = [dotted,rounded corners,red]

\begin{figure}[t]
\centering
\resizebox{0.9\linewidth}{!}{
\begin{tikzpicture}
\node[block](ModeAuto) at (-1.5, 0) {Automatic};
\node[block](ModeInter) at (3.5, 0) {Interactive};
\draw(ModeAuto) -- (-1.5, -1);
\draw(ModeInter) -- (3.5, -1);

\draw[group] ($(ModeAuto.north west) + (-0.1,+0.1)$) rectangle ($(ModeInter.south east) + (0.1,-0.1)$);
\node[pill](1) at (-5.7, 0) {1};
\draw[group] (1) -- ($(ModeAuto.north west) + (-0.1,-0.3)$);

\node[block](Parser) at (-4, -1) {Parser};
\node[block](State) at (6, -1) {State};
\draw(Parser) -- (-1.5, -1);
\draw(State) -- (3.5, -1);
\node[block, minimum height=3em](Loop) at (-1.5, -2.3) {Main Loop};
\draw(-1.5, -1) -- (Loop);
\draw(Loop) -- (1, -2.3);
\node[block, minimum height=3em](Calculus) at (1, -2.5) {Calculus};
\draw(1, -1) -- (Calculus);
\node[block, minimum width=20em](Commands) at (1, -1) {Commands};
\draw(Calculus) -- (3.5, -2.5);

\draw[group] ($(Parser.north west) + (-0.1,+0.1)$) rectangle ($(State.south east) + (0.1,-0.1)$);
\node[pill](2) at (-5.7, -1) {2};
\draw[group] (2) -- ($(Parser.north west) + (-0.1,-0.35)$);

\node[block, minimum height=3em](Translation) at (3.5, -2.7) {Translation};

\draw[group] ($(Loop.north west) + (-0.1,+0.1)$) rectangle ($(Translation.south east) + (0.1,-0.1)$);
\node[pill](3) at (-5.7, -2.5) {3};
\draw[group] (3) -- ($(Loop.north west) + (-0.1,-0.7)$);

\node[block, minimum width=20em](Basic) at (1, -4) {Basic processing \& utilities};
\draw(Loop) -- (Basic);
\draw(Calculus) -- (Basic);
\draw(Translation) -- (Basic);
\node[block, minimum width=20em](Rep) at (1, -5) {Representation};
\draw(Basic) -- (Rep);

\draw[group] ($(Basic.north west) + (-0.1,+0.1)$) rectangle ($(Rep.south east) + (0.1,-0.1)$);
\node[pill](4) at (-5.7, -4.5) {4};
\draw[group] (4) -- ($(Basic.west) + (-0.1,-0.5)$);
\end{tikzpicture}
}
\caption{\Leo{}'s architecture}
\label{fig:architecture} \vspace*{-.5em}
\end{figure}

\section{Improvements}
The TPTP problem set is the canonical benchmark by which theorem
provers are presently evaluated. We
accompany the description of improvements in this section with TPTP
problem names whose solution is affected by the feature. These
problems consist of THF problems drawn from TPTP 5.4.0. We have used E
version 1.6 as the backend ATP. Our tests were run on a 2GHz AMD
Opteron with 4GB RAM, and given 60-second timeout.  LEO-II was
compiled with OCaml 3.11.2.

\subsection{ATP interface} \label{sec:atpInterface}
\Leo{} cooperates with other provers in order to maximise its
potential. We improved \Leo's translation to FOL
in recognition of this. Version \version\ includes a better
translation into FOF, an experimental translation into TFF \cite{sutcliffe2012tptp}, and
supports additional backend ATPs.

\paragraph{Translation into FOL.}
Alongside the old translations which were previously implemented in
\Leo, version \version\ features a new translation module which was
written from scratch. This module contains an intermediate language to
which problems are first translated, before being transformed further
and printed into a specific target syntax. HOL-to-FOL translations
consist of a pipeline of functions which bring HOL formulas into this
intermediate language, applying analyses and transformations along the
way. We are also experimenting with lighter encoding of type information.
We have closely followed Claessen et al~\cite{Claessen:2011:SOM:2032266.2032283}
to implement their monotonicity analysis by producing a SAT encoding,
which we send to MiniSat using an interface adapted from Satallax~\cite{backes2011analytic}.

\Leo's old and new FOF encodings can be used via the command-line
arguments \texttt{--translation fully-typed} and
\texttt{--translation fof\_full} respectively.
The gain of \texttt{fof\_full} over
\texttt{fully-typed} is due to improved handling of formulas---for instance,
the new FOF translation implements full $\lambda$-lifting, which the old
translation didn't. The \texttt{fof\_full} translation is now set as default.
\ignore{
Problems which become provable
using the new translation include GEG003\^1.p, NUM636\^3.p, SEU706\^1.p
and SYO250\^5.p.
}

\paragraph{Backend ATPs.}
\Leo{} is mainly used in combination with E \cite{Schulz02}, and
version \Leo\ \version\ features small improvements in how it interacts
with E. \ignore{Support for SPASS \cite{weidenbach2007system} was added
during past experiments \cite{W37}.} In version \version\ we improved
\Leo{}'s ATP interface and added support for various other backend ATPs,
including remote provers on SystemOnTPTP \cite{sutcliffe2009tptp}.

\subsection{Support for Axiom of Choice}
The default semantics for THF0 is Henkin semantics with choice. Until
version \version, \Leo\ did not support reasoning with choice, unless
na\"ive Skolemization was used---that is, first-order Skolemization
without employing further restrictions (as investigated by Miller
\cite{Miller83}). This enables limited reasoning with choice,
and succeeds in some example cases, but it fails in many
others~\cite[Section 3.2]{C17}.

In order to extend \Leo\ to support the axiom of choice (AC),
instances of AC could be automatically added to the input problem.
An example is the following instance of AC for type $(\iota\rightarrow o)\rightarrow\iota$:
\begin{equation}\label{example:ac}
\exists E_{(\iota\rightarrow o)\rightarrow
  \iota} \forall P_{(\iota\rightarrow o)}. \: \exists X_\iota (P \: X)
\Rightarrow P \: (E \: P)
\end{equation}
However, such
kinds of impredicative axioms should generally be avoided in automated
proof search since they allow for simulation of the
cut rule in any Henkin-complete THF prover~\cite{J18}.

Our approach involves adding two new rules to \Leo: \rl{detectChoiceFn}
and \rl{choice}.
The first rule detects and removes instances of AC, such as (\ref{example:ac}) above,
and keeps a register of choice functions $\ChoiceFuns$.
$\ChoiceFuns$ always contains at least one choice function symbol
for each choice type.
The second rule gives the semantics to choice functions.
Taken together, these rules allow AC-valid reasoning without the
risk of cut-simulation.

In more detail, rule \rl{detectChoiceFn} removes choice-axiom
clauses from the search space and registers the corresponding choice
function symbols $f$ in $\ChoiceFuns$. 
\begin{mathpar}
\begin{prooftree}
[P X]^\false \vee [P (f_{(\alpha\rightarrow o)\rightarrow \alpha} P)]^\true
\justifies
 \ChoiceFuns \longleftarrow \ChoiceFuns\cup\{f_{(\alpha\rightarrow o)\rightarrow \alpha}\}
\thickness=0.08em
\using{\rl{detectChoiceFn}}
\label{detectChoiceFn}
\end{prooftree}
\end{mathpar}
Rule \rl{choice} investigates whether a term
$\epsilon_{(\alpha\rightarrow o)\rightarrow \alpha}
\mathbf{B}_{\alpha\rightarrow o}$ (where $\epsilon\in\ChoiceFuns$ is a
registered choice function or a free variable) is contained as a
subterm of a literal $[\mathbf{A}]^p$ in a clause $C$. In this case it
adds the instantiation of AC at type $(\alpha\rightarrow o)\rightarrow
\alpha$, and with term $\mathbf{B}$, to the search space. Side-conditions
guard against unsound reasoning, such as the `uncapturing' of free
variables in $\mathbf{B}$:
\begin{mathpar}
\begin{prooftree}
\begin{array}{c}
\\
C:= \mathbf{C'} \vee [\mathbf{A}[E_{(\alpha\rightarrow o)\rightarrow \alpha} \mathbf{B}]]^p
\end{array}
 \quad
\begin{array}{c}
\epsilon\in\ChoiceFuns, \; E = \epsilon \textit{ or } E \in \textit{freeVars(C)},
\\ \textit{freeVars}(\mathbf{B})\subseteq \textit{freeVars}(C), Y\,\textit{fresh}
\end{array}
\justifies
  [\mathbf{B} \; Y]^\false \vee [\mathbf{B} \; (\epsilon_{(\alpha\rightarrow o)\rightarrow \alpha}
  \mathbf{B})]^\true
\thickness=0.08em
\using{\rl{choice}}
\label{choice}
\end{prooftree}
\end{mathpar}
%
%
%
%
Rules \rl{detectChoiceFn} and \rl{choice} are obviously sound:
\rl{detectChoiceFn} simply removes clauses from the search space, and for any choice function $f$, the rule
\rl{choice} only introduces new instances of the corresponding choice
axiom. 

There is a correspondence with the handling of choice in Satallax.
Satallax too considers only selective
instantiations of AC in order to avoid
cut-simulation.
For instance, when~(\ref{example:ac}) is assumed, the terms $\mathbf{T}$ which Satallax considers to be eligible
instantiations for variable $P$ are those occurring in formulas
of the following forms in a tableau branch (and where $\epsilon$ is a
choice function):
$(\epsilon \; \mathbf{T}) \; \mathbf{S_1} \: \ldots \: \mathbf{S_n}$ or
$\neg ((\epsilon \; \mathbf{T}) \; \mathbf{S_1} \: \ldots \: \mathbf{S_n})$,
or the disequations
$(\epsilon \; \mathbf{T}) \: \mathbf{S_1} \: \ldots \: \mathbf{S_n} \not= \mathbf{S}$ or
$\mathbf{S} \not= (\epsilon \; \mathbf{T}) \: \mathbf{S_1} \: \ldots \: \mathbf{S_n}$.
It is easy to see that our rule \rl{choice}, which is less
restrictive, subsumes these cases. We also experimented with
Satallax's approach in \Leo\ but this led to worse results.
Our choice rule is more closely related to that of Mints~\cite{Mints99JSL}.
Use of the choice rules can be disabled using the \verb|-nuc| command-line
switch.

\subsection{Detection of defined equality}
\emph{Primitive equality} in HOL refers to the use of the interpreted constant `='.
Equality can also be \emph{defined} in HOL---for example, as $\lambda X_\alpha
\lambda Y_\alpha \forall P_{\alpha\rightarrow o}. \; P \; X \Rightarrow P
\; Y$ or $\lambda X_\alpha \lambda Y_\alpha \forall
Q_{\alpha\rightarrow\alpha\rightarrow o}. \; \forall Z_\alpha (Q \; Z
\; Z) \Rightarrow Q \; X \; Y$. The former is known as Leibniz
equality and the latter we call Andrews equality (cf.~\cite{Andrews2002}, Exercise X5303). 
Both Leibniz and Andrews equality support
cut-simulation due to their impredicative nature~\cite{J18}, and should thus be avoided in proof
automation.  In fact, using primitive, rather than defined, equality may save many \emph{primitive
substitution} steps in proofs. Such steps involve instantiations of set variables, and this generally involves blind guessing.
Examples of the benefit of using primitive, rather than defined, equality have been given in the literature~\cite[Sections 5.1
and 5.2]{C17}. In order to address this issue we added the
following two rules to \Leo's calculus; they instantiate the set variable $P$ with primitive equality:
\begin{mathpar}
\begin{prooftree}
\mathbf{C} \; \vee \; [P \; \mathbf{A}]^\false \; \vee \; [P \; \mathbf{B}]^\true
\justifies
\mathbf{C}\{\lambda X. \; \mathbf{A} = X/P\} \vee [\mathbf{A} = \mathbf{B}]^\true
\thickness=0.08em
\using{\rl{LeibEQ}}
\label{LeibEQ}
\end{prooftree}
\quad
\begin{prooftree}
\mathbf{C} \; \vee \; [P \; \mathbf{A} \; \mathbf{A}]^\false
\justifies
\mathbf{C}\{\lambda X \lambda Y. \; X = Y/P\}
\thickness=0.08em
\using{\rl{AndrEQ}}
\label{AndrEQ}
\end{prooftree}
\end{mathpar}
Soundness of \rl{LeibEQ} and \rl{AndrEQ} is obvious, since both rules
simply realise specific instances of primitive substitution. 
For improved configurability, either rule can be individually disabled from the command-line
by using the switches \verb|-nrleq| and \verb|-nraeq| respectively.
If \rl{LeibEQ} is used in combination with the new FOF translations
(see Section~\ref{sec:atpInterface}) several TPTP problems whose
previous SZS~\cite{sutcliffe2008szs} status was `Unknown' can now be
solved by \Leo. Examples include SYO246\^{}5.p,
SYO244\^{}5.p, NUM817\^{}5.p, NUM816\^{}5.p, and NUM814\^{}5.p.  There
are also many problems that can now be solved with primitive
substitution (blind guessing) disabled when \rl{LeibEQ} and
\rl{AndrEQ} are available. 
Overall, these two new rules lead to significantly
better coverage using the lighter primitive-substitution search modes
\verb|-ps 0| or \verb|-ps 1|.

\subsection{Strategy scheduling}
Strategy schedules were added to \Leo{} in version 1.2 and the
catalogue of schedules has slowly increased in the versions that followed.
In version \version\ we recoded the strategy-scheduling feature to facilitate
the encoding of new strategies, to improve code reuse with other parts of \Leo, and
to have greater flexibility when encoding strategies.
\ignore{The new setup affords greater flexibility: for example, the
new setup can schedule varying number of strategies (depending on the
problem being processed) and each schedule could be of varying duration.
This has opened up many opportunities for experimentation and tuning.}

We are also interested in computing strategies on-the-fly based on
problem characteristics, and version \version\ carries out some small
initial checks (e.g. size of the problem, and whether it contains
instances of AC), and schedules strategies based
on that limited analysis. Optimising this further remains as future
work.

\subsection{Other improvements}
Numerous other additions were made to \Leo{}.
Previously, \Leo\ was entirely focused on refutation: that is, until version
\version, in terms of the SZS classification,
\Leo\ would judge a problem to be a Theorem (if a refutation exists),
Unsatisfiable (if the problem's axioms themselves can be refuted), or
diverge (by extending the preunification depth and reattempting a
refutation).  It can now classify Satisfiable problems and detect
CounterSatisfiable problems, thus improving both \Leo's precision and
termination behaviour. The added support for choice was very relevant for achieving this.

\Leo's unification algorithm has been redone, and can be set (from
the command-line) to disregard Boolean and functional extensionality.
This has strengthened \Leo's behaviour in non-extensional problems,
since disabling the extensional behaviour shrinks the search space.

Numerous other improvements and fixes have been made: these range from
system features (such as the parser, status reporting, avoiding
redundant computations, etc) to deeper areas in the calculus and main
loop (including factorisation, subsumption, and clause selection).

\section{Future work}

We have started experimenting with using term orderings to influence
literal selection.  We also plan to revise \Leo{}'s internals to make
full use of the potential benefit they offer. For instance, the shared
term graph is currently underutilised.

\ignore{More work is needed to compute better schedules.
There is plenty of room for more experimentation and optimisation here.
For instance, scheduling can be refined to detail the invocations of
the backend ATP, such as the duration and frequency of these
invocations. Experiments have shown that this can have a significant
influence on results. For instance, in truly higher-order problem,
which requires more work from \Leo{}'s calculus, the first dispatches
to a first-order ATP are usually wasted time.}
More work is needed to compute better schedules, paired with better problem
analyses. Such analyses can determine the
scheduling of specific strategies, which can be better tuned to the
problem.

The ATP interface can be improved further to call multiple backend ATPs in parallel.
Experiments comparing 30-second invocations of \Leo\, on all THF
problems, supported by provers E (version 1.6), SPASS (version 3.5) \cite{SPASS}
and Vampire (version 2.6) \cite{VAMPIRE} showed us that there were 37, 5 and 20
theorems that were proved exclusively by \Leo(E), \Leo(SPASS) and
\Leo(Vampire), respectively. And there were 31, 95 and 98 theorems
that \Leo(E), \Leo(SPASS) and \Leo(Vampire) missed, but which one of
the others could prove.

\ignore{Currently, \Leo{}
invokes a single backend ATP and blocks until the result is received.
It would be far better to adopt a different model, where \Leo{} can
continue running its main loop in parallel to the ATP dispatch.
Furthermore, \Leo{} could start multiple backend ATPs. Similar
techniques are already used by tools such as
Sledgehammer~\cite{paulson2010three} and Why3~\cite{bobot2011why3}.}
Supporting various ATP backends increases the scope for peephole
optimisation; we have not yet investigated this.
The translation module can be optimised further, and extended to
target more formats. Table~\ref{tab:comparison:fol} one shows how
the new HOL-to-FOL translation (\texttt{fof\_full}) and its lighter
variant (\texttt{fof\_experiment}) are superior to \Leo's preexisting
encoding (\texttt{fully\_typed}). In future work we plan to improve
\texttt{fof\_experiment} further and make it the default translation.


\begin{table}[t]
\centering
\begin{tabular}{c@{\hskip 3mm}c@{\hskip 3mm}c@{\hskip 3mm}c}
\textbf{SZS Status} & \texttt{fully-typed} & \texttt{fof\_full} & \texttt{fof\_experiment} \\ \hline
\textbf{Thm} & \ignore{1609}64.8 & \ignore{1611}64.9 & \ignore{1622}65.3\\
\textbf{All} & \ignore{1736}60.9 & \ignore{1738}61 & \ignore{1747}61.3
\end{tabular}
\caption{Comparing FOL encodings in \Leo\ \version\ (30s timeout).
Table shows the percentage of matches between \Leo's SZS output and the `Status' field of problems.}
\label{tab:comparison:fol}\vspace*{-1em}
\end{table}

\section{Conclusion}
Version \version\ of \Leo\ includes various improvements which affect
its performance and completeness.  To obtain a broader picture, we
compared the results of using \Leo\ version \version\ with earlier
versions, and the results are shown in Table~\ref{tab:comparison}.  In
this experiment we counted the matches between \Leo's SZS output and
the TPTP problem's SZS status (included in its header).\footnote{This
  also means that `Unknown' problems which \Leo\ now classifies as
  `Theorem' count against us, but this experiment was only intended to
  offer a rough idea of progress.} All the net gains are positive, but
a more thorough evaluation (on different benchmarks, and considering
various parameters) remains as future work.  Within a 30s timeout,
\Leo\ version \version\ can classify 196 more problems than its
predecessor. The main boost in this version is provided by the
detection of non-theorems ($\frac{125}{196}$).


\setlength{\tabcolsep}{8pt} 
\begin{table}[t]
\centering
\begin{tabular}{ccccccc}
\textbf{Timeout} (s) & \multicolumn{2}{c}{\textbf{v1.2}} & \multicolumn{2}{c}{\textbf{v1.4.3}} & \multicolumn{2}{c}{\textbf{v\version}} \\
   & \textit{Thm} & \textit{All} & \textit{Thm} & \textit{All} & \textit{Thm} & \textit{All} \\ \hline
30 & \ignore{1451}58.4 & \ignore{1456}51.1 & \ignore{1542}62.1 & \ignore{1551}54.4 & \ignore{1622}64.3 & \ignore{1747}61.3 \\
60 & \ignore{1458}58.7 & \ignore{1463}51.3 & \ignore{1613}65 & \ignore{1622}56.9 & \ignore{1666}67.1 & \ignore{1793}62.9 \\
\end{tabular}
\caption{Percentage match between different versions of \Leo\ and the Status field of TPTP problems. LEO-II version 1.2 was the winner of the CASC competition in 2010, and version 1.4.3 was the last public release. Version \version\ was run with the \texttt{fof\_experiment} encoding.}
\label{tab:comparison}\vspace*{-1em}
\end{table}

{\small
\subsubsection*{Acknowledgements:}
Work on \Leo{} was supported by the UK Engineering and Physical
Sciences Research Council (grant number EP/D070511/1), and the German
Research Foundation under grant BE 2501/9-1.
The second author was supported by a grant from the German Academic
Exchange Service (DAAD) during a visit to Freie Universit\"at Berlin,
during which the work described in this paper was carried out.
We thank Chad Brown for allowing us to use Satallax's MiniSat
interface, and thank Mike Gordon, Grant Passmore, and Larry Paulson
for feedback on this article.
}


\begin{thebibliography}{10}

\bibitem{Andrews2002}
P.B. Andrews.
\newblock {\em An Introduction to Mathematical Logic and Type Theory: To Truth
  Through Proof}.
\newblock Applied Logic Series. Springer, 2002.

\bibitem{backes2011analytic}
J.~Backes and C.E. Brown.
\newblock Analytic tableaux for higher-order logic with choice.
\newblock {\em Journal of Automated Reasoning}, 47(4):451--479, 2011.

\bibitem{J5}
C. Benzm{\"u}ller.
\newblock Comparing approaches to resolution based higher-order theorem
  proving.
\newblock {\em Synthese}, 133(1-2):203--235, 2002.

\bibitem{J18}
C. Benzm{\"u}ller, C.E. Brown, and M. Kohlhase.
\newblock Cut-simulation and impredicativity.
\newblock {\em Logical Methods in Computer Science}, 5(1:6):1--21, 2009.

\bibitem{C17}
C. Benzm{\"u}ller and C.E. Brown.
\newblock {A Structured Set of Higher-Order Problems}.
\newblock {\em Proc.~of TPHOLs 2005}, number 3603 in LNCS, pp.
  66--81. Springer, 2005.

\bibitem{C26}
C. Benzm{\"u}ller, F. Theiss, L. Paulson, and A. Fietzke.
\newblock {LEO-II} - a cooperative automatic theorem prover for higher-order
  logic.
\newblock {\em Proc. of IJCAR 2008}, vol. 5195 of {\em LNCS}, pp.
  162--170. Springer, 2008.

\bibitem{Claessen:2011:SOM:2032266.2032283}
K. Claessen, A. Lilliestr\"{o}m, and N. Smallbone.
\newblock Sort it out with monotonicity: translating between many-sorted and
  unsorted first-order logic.
\newblock {\em Proc. of CADE 2011}, vol. 6803 of {\em LNCS}, pp. 207--221. Springer, 2011.

\bibitem{Miller83}
D.A. Miller.
\newblock {\em Proofs in Higher-Order Logic}.
\newblock PhD thesis, Carnegie Mellon U., 1983.


\bibitem{Mints99JSL}
G.~Mints.
\newblock Cut-elimination for simple type theory with an axiom of choice.
\newblock {\em Journal of Symbolic Logic}, 64(2):479-485, 1999.

\bibitem{VAMPIRE}
A. Riazanow and A. Voronkov.
\newblock The design and implementation of {VAMPIRE}.
\newblock {\em AI Commun.} 15(2):91--110, 2002.

\bibitem{Schulz02}
S. Schulz.
\newblock {E -- A Brainiac Theorem Prover}.
\newblock {\em AI Commun.}, 15(2/3):111--126, 2002.

\bibitem{W47}
N. Sultana and C. Benzm{\"u}ller.
\newblock Understanding {LEO-II's} proofs.
\newblock {\em The 9th International Workshop on the Implementation of Logics
  (IWIL-2012, affiliated with LPAR-2012)}, Merida, Venezuela, 2012.

\bibitem{sutcliffe2008szs}
G.~Sutcliffe.
\newblock {The SZS ontologies for automated reasoning software}.
\newblock {\em Proc. of the LPAR Workshops: Knowledge Exchange:
  Automated Provers and Proof Assistants, and The 7th International Workshop on
  the Implementation of Logics}, vol. 418, pp. 38--49. CEUR Workshop
  Proc., 2008.

\bibitem{sutcliffe2012tptp}
G.~Sutcliffe, S.~Schulz, K.~Claessen, and P.~Baumgartner.
\newblock The TPTP typed first-order form with arithmetic.
\newblock {\em Proc. of LPAR 2012}, 
vol. 7180 of {\em LNCS}, pp. 406--419. Springer, 2012.

\bibitem{sutcliffe2009tptp}
G. Sutcliffe.
\newblock {The TPTP problem library and associated infrastructure}.
\newblock {\em Journal of Automated Reasoning}, 43(4):337--362, 2009.

\bibitem{sutcliffe2010automated}
G. Sutcliffe and C. Benzm\"{u}ller.
\newblock {Automated Reasoning in Higher-Order Logic using the TPTP THF
  Infrastructure}.
\newblock {\em Journal of Formalized Reasoning}, 3(1):1, 2010.

\bibitem{SPASS}
C. Weidenbach, D. Dimova, A. Fietzke, R. Kumar, M. Suda and P. Wischnewski.
\newblock SPASS Version 3.5. 
\newblock {\em Proc. of CADE 2009}, vol. 5663 of LNCS, pp. 140--145, Springer.
\end{thebibliography}
\bibliographystyle{plain}

\end{document}